\documentclass[slac_one]{revtex4}
\usepackage{graphicx}
\usepackage{fancyhdr}
\usepackage{tabularx}
\makeatletter
\newcommand{\figcaption}[1]{\def\@captype{figure}\caption{#1}}
\newcommand{\tblcaption}[1]{\def\@captype{table}\caption{#1}}
\makeatother
\begin{document}
\title{Recent Results from  KamLAND} 

%

\author{Koichi Ichimura for the KamLAND Collaboration}
\affiliation{University of California, Berkeley,  California 94720, USA}

\begin{abstract}
The main goal of the KamLAND reactor $\bar{\nu}_{e}$ experiment is a search for $\bar{\nu}_{e}$ oscillation using  inverse-$\beta$ decay reaction in  1,000 ton of ultra-pure liquid scintillator.  The data-set is 1490.8 days from Mar. 2002 to May 2007. The best-fit oscillation parameters are $\Delta$m$^{2}_{21}$ =  7.58$^{+0.14}_{-0.13}$ ($stat.$) $\pm$ 0.15 ($syst.$) $\times$ 10$^{-5}$ eV$^{2}$ and tan$^{2}\theta_{12}$ = 0.56$^{+0.10}_{-0.07}$ ($stat.$) $^{+0.10}_{-0.06}$ ($syst.$). The statistical significance for reactor  $\bar{\nu}_{e}$ disappearance is  8.8$\sigma$,  and an  undistorted $\bar{\nu}_{e}$ energy spectrum is disfavored at $>$ 5$\sigma$.   
\end{abstract}

\maketitle

\thispagestyle{fancy}


\section{Introduction} 
KamLAND(\textbf{Kam}ioka \textbf{L}iquid scintillator \textbf{A}nti-\textbf{N}eutrino \textbf{D}etector)
investigates neutrino oscillation parameters by observing electron anti-neutrinos (${\bar{\nu}_{e}}$) from distant nuclear reactors.  Previously, KamLAND  reported the first evidence of ${\bar{\nu}_{e}}$ disappearance~\cite{KamLAND_1st} and  evidence of spectral distortion of the reactor ${\bar{\nu}_{e}}$ energy spectrum~\cite{KamLAND_2nd}. Furthermore, KamLAND published the first investigation of geologically produced anti-neutrinos (geo-neutrinos) from radioactive decay in the Earth\cite{KamLAND_Geo}. 

In our recent analysis~\cite{KamLAND_3rd},  we have enlarged the  fiducial volume radius from 5.5m to 6.0m, extended the  analysis threshold down to  the inverse-$\beta$ decay energy threshold and reduced the systematic uncertainties  in the number of target protons and the background. 

\section{Detection method and analysis improvements}
\subsection{Detection method}
 ${\bar{\nu}_{e}}$  are detected via  the inverse $\beta$ decay reaction, $\bar{\nu}_{e}$ + p $\rightarrow$ n + e$^{+}$.  The emitted positron annihilates with an electron and emits two 0.511MeV $\gamma$-rays. This signal appears as a prompt signal and its energy consist of the positron kinetic energy and these two $\gamma$-rays. The prompt scintillation light from the e$^{+}$ gives a measure of the ${\bar{\nu}_{e}}$ energy,  $E_{\bar{\nu}_{e}} \simeq E_{prompt} + \bar{E_{n}}$ + 0.8MeV, where $\bar{E_{n}}$ is the average neutron recoil energy, O(10keV).  The neutron thermalizes by elastic scattering in the liquid scintillator and captures on a  proton after  207.5 $\pm$ 2.8$\mu$sec,   creating a deuteron with 2.2MeV $\gamma$ emission. The neutron capture is  the delayed signal.  By using the delayed coincidence method, the backgrounds that  do not  have timing or space correlations are significantly reduced. 
\subsection{Off-axis calibration}
We recently used an off-axis calibration system to investigate the vertex position reconstruction performance.  Up to 5.5m radius, the vertex reconstruction deviation is within 3cm  at any (R, $\theta$, $\phi$), corresponding to 1.6$\%$ fiducial volume uncertainty.   $^{12}$B/$^{12}$N spallation products are used to extrapolate this uncertainty up to 6.0m fiducial radius, resulting in 1.8$\%$ uncertainty.  The vertex and energy resolution are also estimated from  off-axis and z-axis calibration data. The estimated vertex resolution is $\sim$ 12cm/$\sqrt{E(\mathrm{MeV})}$ and the energy resolution is 6.5$\%$/$\sqrt{E(\mathrm{MeV})}$. 

\subsection{Likelihood selection}
In Ref.\cite{KamLAND_2nd}, 5.5m fiducial radius and  2.6MeV prompt energy threshold were applied to the reactor ${\bar{\nu}_{e}}$ analysis.   To get more statistics and to use  the full reactor energy spectrum, a larger fiducial volume  and a lower energy threshold are required.  However, accidental coincidences due to $^{40}$K, $^{210}$Bi and $^{208}$Tl increase near the balloon surface (R = 6.5m)  reducing the signal-noise ratio.    To maintain high efficiency for  ${\bar{\nu}_{e}}$  detection, we apply a likelihood selection using event characteristics. Firstly we construct a probability density function (PDF) for accidental coincidence events,  $f_{acci}(E_{p}, E_{d}, \Delta R, \Delta T, R_{p}, R_{d} )$ by using an off-time (10ms to 20 sec)  window. A PDF for the ${\bar{\nu}_{e}}$, $f_{\bar{\nu}_{e}}(E_{p}, E_{d}, \Delta R, \Delta T, R_{p}, R_{d} )$ is constructed from a MC simulation. For the $E_{prompt}$ distribution, we select an no-oscillation reactor spectrum including a contribution from geo-neutrinos. A discriminator value, $L = \frac{f_{\bar{\nu}_{e}}}{f_{acci} + f_{\bar{\nu}_{e}}}$ is calculated for all candidate pairs that pass the primary cuts.  We choose a selection value $L_{i}^{cut}$ in $E_{p}$ bins of 0.1 MeV, where  $L_{i}^{cut}$ is the value which has the maximum figure-of-merit.  Table \ref{tbl:uncertainty} shows the summary of  the systematic uncertainties in the analysis. The dominant uncertainties come from the reactor-related ones.

\begin{center}
\begin{table}[h]
\caption{Estimated systematic uncertainties relevant for the neutrino oscillation parameters.} \label{tbl:uncertainty}
\begin{tabular}{c|cccc}  \hline \hline
& Detector-related($\%$)& & Reactor-related($\%$)&\\ \hline
$\Delta$m$_{21}^{2}$ & Energy scale & 1.9 & $\bar{\nu}_{e}$ spectra &  0.6 \\ \hline
& Fiducial Volume & 1.8 &  $\bar{\nu}_{e}$ spectra &  2.4 \\
Event rate & Energy threshold & 1.5 &  Reactor power &  2.1 \\
& Efficiency & 0.6 &  Fuel composition &  1.0 \\
& Cross section & 0.2 &  Long-lived nuclei &  0.3 \\ \hline \hline
\end{tabular}
\end{table}
\end{center}

\subsection{Background}
\begin{figure}[h]
 \def\@captype{table}
\begin{minipage}[c]{.44\textwidth}
\tblcaption{Summary of estimated backgrounds} 
 \begin{center}
\begin{tabular}{c|c}  \hline
 Background&contribution\\ \hline
Accidentals & 80.5 $\pm$ 0.1 \\
$^{8}$He/$^{9}$Li & 13.6 $\pm$ 1.0 \\
Fast neutrons and Atmospheric $\nu$ & $<$ 9.0 \\
$^{13}$C($\alpha$, $n$)$^{16}$O G.S. & 157.2 $\pm$ 17.3 \\
$^{13}$C($\alpha$, $n$)$^{16}$O $^{12}$C($n$, $n\gamma$)$^{12}$C (4.4 MeV $\gamma$) & 6.1 $\pm$ 0.7 \\
$^{13}$C($\alpha$, $n$)$^{16}$O 1st excited state (6.05 MeV $e^{+}e^{-}$) & 15.2 $\pm$ 3.5 \\
$^{13}$C($\alpha$, $n$)$^{16}$O 2nd excited state (6.13 MeV $\gamma$) & 3.5 $\pm$ 0.2 \\ \hline
Total & 276.1 $\pm$ 23.5 \\ \hline
\end{tabular}
\end{center}
 \label{tbl:BG}
\end{minipage}
\hfill
  \begin{minipage}[c]{.44\textwidth}
  \begin{center}
  \includegraphics[width=60mm,angle=270]{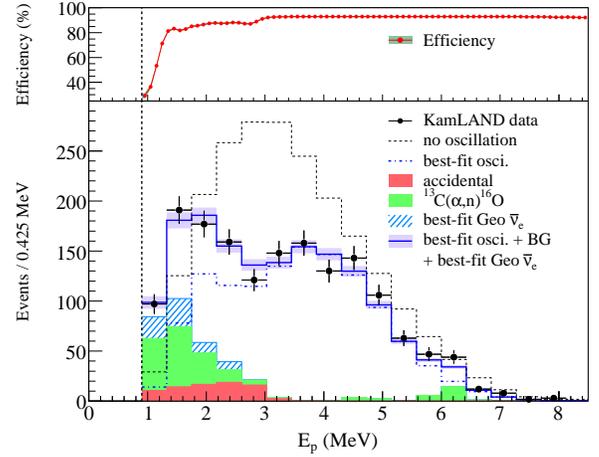}
\caption{Prompt energy spectrum of $\bar{\nu}_{e}$ candidate events. All histograms corresponding to reactor spectra and expected backgrounds incorporate the energy-dependent selection efficiency(top). The shaded background and geo-neutrino histograms are cumulative. Statistical uncertainties are shown for the data. The band on the blue histogram indicates the event rate systematic uncertainty.} \label{fig:JACpic2-f1}
\end{center}
\end{minipage}
\end{figure}

The dominant background is caused by the $^{13}$C($\alpha$,$n$)$^{16}$O reaction. The primary $\alpha$ source is the decay of $^{210}$Po, a decay-daughter of  $^{222}$Rn  introduced during the  liquid scintillator filling. In Ref.\cite{KamLAND_2nd}, a 32\% error  for the reaction to $^{16}$O ground state was assigned from uncertainties of  $^{210}$Po rate, the cross section to $^{16}$O ground state, neutron angular distribution and proton quenching. A conservative  100\% error  was assigned to  the excited states since the cross section was not well known.  To measure the ($\alpha$,$n$) prompt spectrum, a detector calibration with a  $^{210}$Po$^{13}$C source was performed.  The expected spectrum was  also simulated by the $^{13}$C($\alpha$, $n$)$^{16}$O  cross sections from Ref.\cite{CS}\cite{CS2} and compared with the data. 
 The new cross section measurement\cite{CS} has a 4\% uncertainty in the total cross section,  but it does not specify  the final state separately. Thus the subtraction of  the reaction rate of the excited states estimated from the cross section in Ref.\cite{CS2}   is needed for  this new measurement.  
The first $^{16}$O excited state emits $e^{+}e^{-}$ , but they cannot escape from the source capsule . In KamLAND, these events are observed at $\sim$ 1.02MeV from the annihilation gammas.
Therefore, the reaction rate of each state can be measured separately by $^{210}$Po$^{13}$C calibration source.
 The rate of  the second excited state  was consistent with the calculation\cite{CS2} without scaling. For the first excited state, the calculated spectrum agreed with the calibration data if  scaled by 0.6. After subtraction of the excited state cross sections,  the ground state was obtained.  The reaction rate of  the ground state  had to be  scaled by 1.05 to get the calculated ground state to be in  good agreement with the data.  
 With these improvements,  we assign an uncertainty of 11\% for the ground state and 20\% for the excited states. Finally, the number of background events due to  $^{13}$C($\alpha$,$n$)$^{16}$O is estimated to be 182 $\pm$ 21.7 events. Table \ref{tbl:BG} shows the summary of the backgrounds.

\section{Results}
 We observe 1609 events in our data-set.   With no ${\bar{\nu}_{e}}$ disappearance, we expect 2179 $\pm$ 89 events from the reactors. The background in this analysis is 276.1 $\pm$ 23.5 events excluding the  geo-neutrino signal.  
 Figure \ref{fig:JACpic2-f1} shows the prompt energy spectrum of selected  ${\bar{\nu}_{e}}$ events and the fitted backgrounds. The unbinned data is assessed with a maximum likelihood fit to two-flavor neutrino oscillation, simultaneously fitting the geo-neutrino contribution. The method uses the absolute time of the event and accounts for time variations in the reactor flux. Earth matter oscillation effects are also included.

\begin{figure*}[htbp]
\begin{tabular}{cc}
\begin{minipage}{.44\textwidth}
\begin{center}
\includegraphics[width=60mm,angle=270]{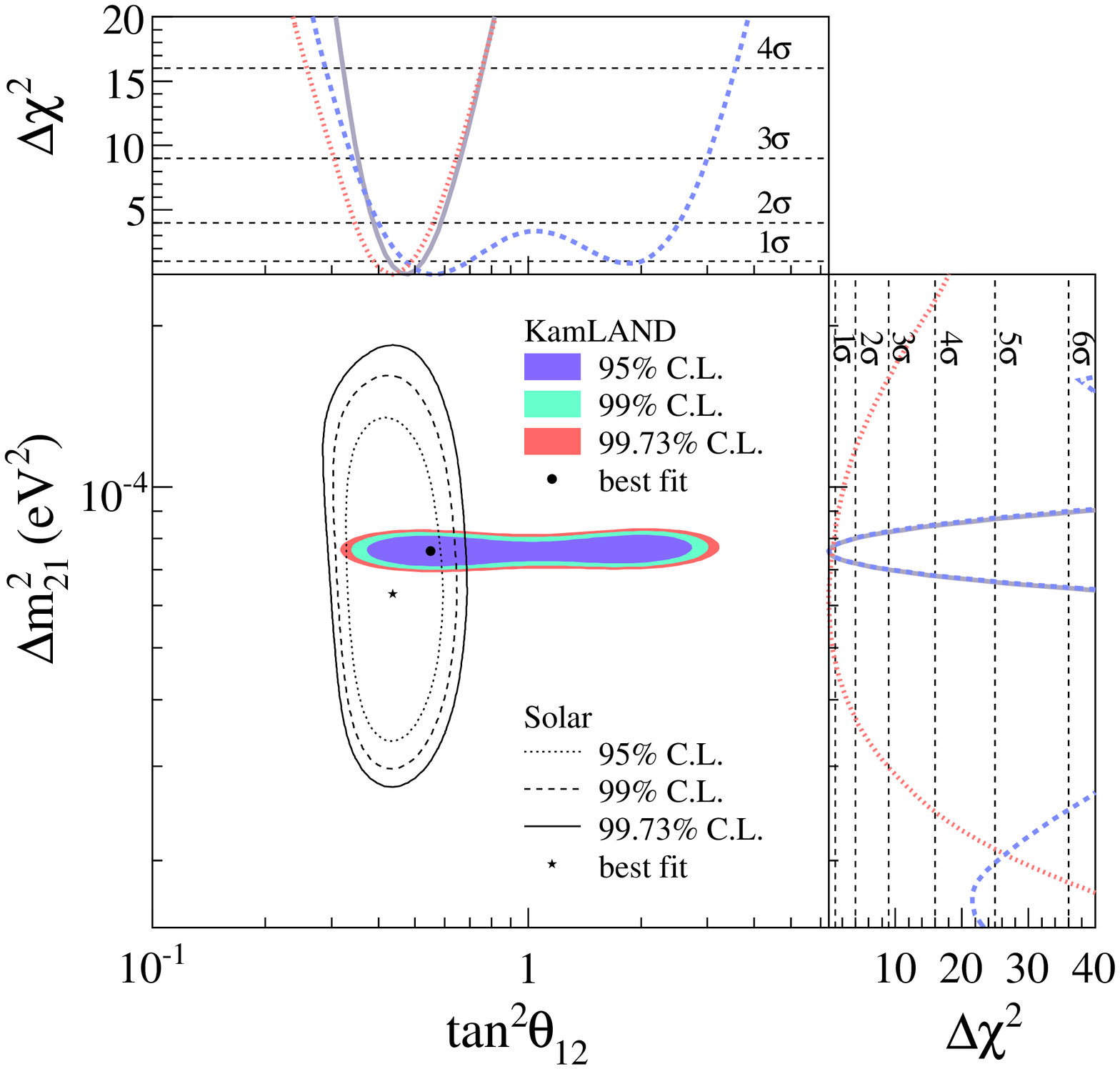}
\caption{Allowed region for neutrino oscillation parameters from KamLAND and solar neutrino experiments. The side panels show the $\Delta\chi^{2}$-profiles for KamLAND(dashed),solar experiments(dotted) and the combination of the two(solid).} \label{fig:JACpic2-f2}
\end{center}
\end{minipage}
\hspace{10mm}
\begin{minipage}{.44\textwidth}
\begin{center}
\includegraphics[width=60mm,angle=270]{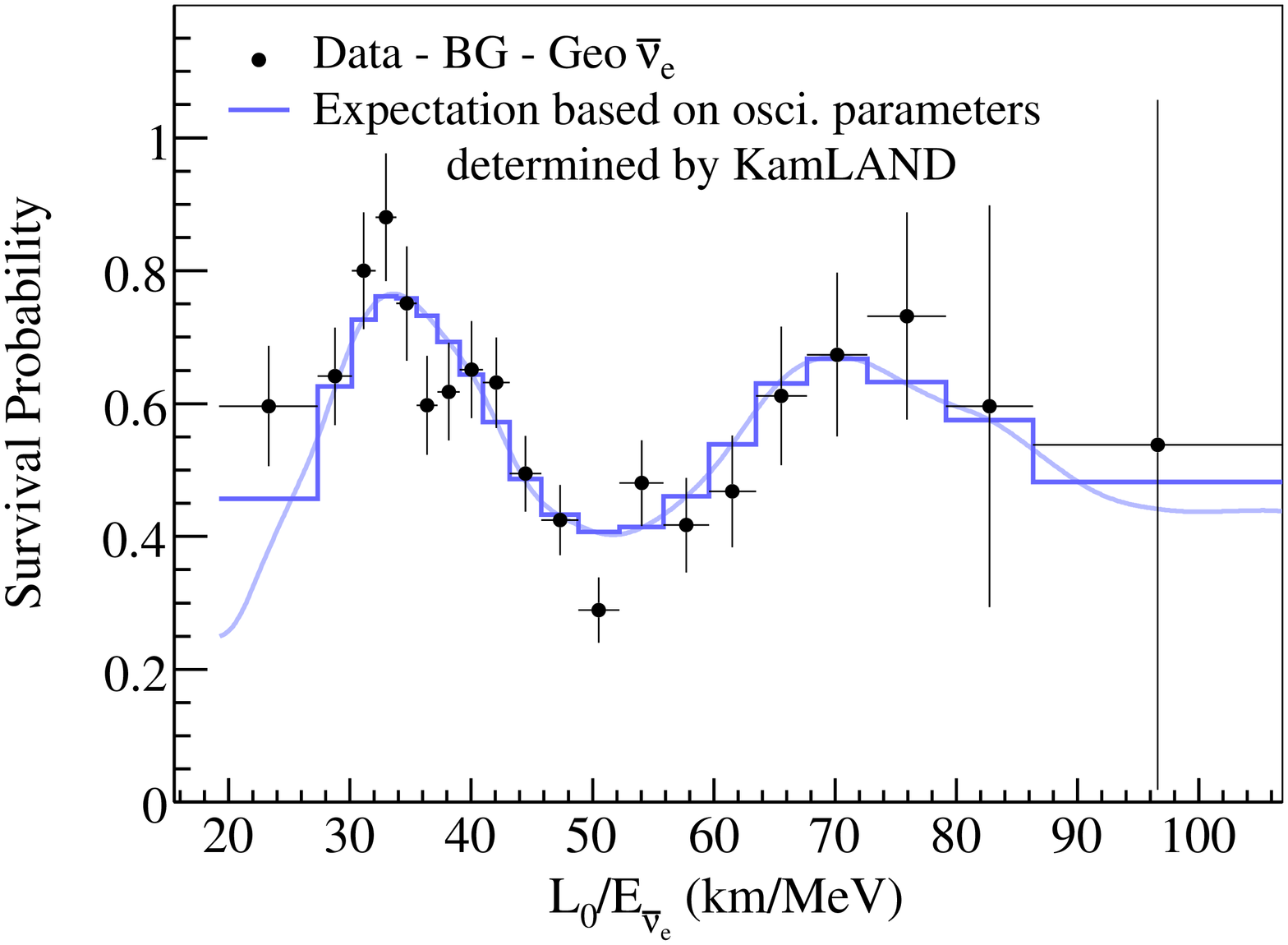}
\caption{Ratio of the background subtracted $\bar{\nu}_{e}$ spectrum to the expectation without oscillation   as a function of L$_{0}$/E. L$_{0}$ is the effective baseline taken as a flux weighted average (L$_{0}$ = 180 km). The energy bins are equal probability bins of the best-fit including all backgrounds. The blue curves are the expectation accounting for the distance to the individual reactors, time dependent flux variations and efficiencies.  The error bar shows the statistical uncertainty only. } \label{fig:JACpic2-f3}
\end{center}
\end{minipage}
\end{tabular} 
\end{figure*}

 The best-fit is shown in Figure \ref{fig:JACpic2-f1}. The best-fit oscillation parameters are $\Delta$m$^{2}_{21}$ =  7.58$^{+0.14}_{-0.13}$ ($stat.$) $\pm$ 0.15 ($syst.$) $\times$ 10$^{-5}$ eV$^{2}$ and tan$^{2}\theta_{12}$ = 0.56$^{+0.10}_{-0.07}$ ($stat.$) $^{+0.10}_{-0.06}$ ($syst.$). A scaled reactor spectrum with no distortion from neutrino oscillation is excluded at more than 5 $\sigma$.  Figure \ref{fig:JACpic2-f2} shows the allowed  contours in the neutrino oscillation parameter space. Only the so-called LMA-I region remains and the LMA-II  and LMA-0 regions are disfavored at $>$ 4$\sigma$. A combined oscillation analysis of KamLAND and  other solar neutrino experiments under the assumption of CPT invariance gives $\Delta$m$^{2}_{21}$ =  7.59  $\pm$ 0.21  $\times$ 10$^{-5}$ eV$^{2}$ and tan$^{2}\theta_{12}$ = 0.47$^{+0.06}_{-0.05}$.  The $\Delta$m$^{2}_{21}$ parameter is strongly determined by the KamLAND experiment.   

  The L$_{0}$ / E distribution is  suitable for illustrating the oscillatory behavior, see  Figure\ref{fig:JACpic2-f3}. In this figure, the data and the best-fit spectrum are divided by the expected spectrum without oscillation.   L$_{0}$ is the flux-weighted effective baseline and is 180 km.  Two oscillation periods are seen, the KamLAND data is in excellent agreement with the oscillation scenario. 
   
  To determine the number of geo-neutrinos, we fit the normalization of the $\bar{\nu}_{e}$ energy spectrum from U and Th decay chains simultaneously with the neutrino oscillation parameters  using the KamLAND and solar data.  Fixing the Th/U mass ratio from planetary data to 3.9, the best-fit value is 73 $\pm$ 27 events. 

\section{Conclusion}
The KamLAND experiment has determined a precise value for the neutrino oscillation parameter  $\Delta$m$^{2}_{21}$ and stringent constraints on $\theta_{12}$ using reactor anti-neutrinos from 55 Japanese reactors over a five years period. The observed neutrino spectrum is distorted and the significance of the spectral distortion is  $>$ 5$\sigma$ and no-spectral distortion is strongly disfavored.  A two-neutrino oscillation analysis of the KamLAND data with 0.9 MeV threshold gives $\tan^{2}\theta_{12}$ = 0.56$^{+0.10}_{-0.07}$ ($stat.$) $^{+0.10}_{-0.06}$ ($syst.$) and $\Delta$m$^{2}_{21}$ = 7.58
$^{+0.14}_{-0.13}$ ($stat.$) $\pm$0.15($syst.$) $\times$ 10$^{-5}$ eV$^{2}$.  The previous result~\cite{KamLAND_2nd}  was $\tan^{2}\theta_{12}$ = 0.46$^{+0.23}_{-0.13}$  and $\Delta$m$^{2}_{21}$ = 7.9
$^{+0.4}_{-0.3}$  $\times$ 10$^{-5}$ eV$^{2}$.  The uncertainties of neutrino oscillation parameters are significantly  reduced and the systematic and statistical uncertainty are now comparable. The LMA-II and LMA-0  which were both allowed  in Ref.~\cite{KamLAND_2nd}  are now disfavored at $>$ 4 $\sigma$ and only  LMA-I  remains. These results are also published in Ref.\cite{KamLAND_3rd}.

\end{document}